\documentclass{article}
\usepackage{amsfonts}
\usepackage{amsmath}

= -2truecm

\begin{document}

\title{Generating entanglement of photon-number states with coherent light via cross-Kerr
nonlinearity}
\author{Zhi-Ming Zhang\thanks{%
Corresponding author: zmzhang@scnu.edu.cn}, Jian Yang, and Yafei Yu \\
Laboratory of Photonic Information Technology,\\
School of Information and Photoelectronics,\\
South China Normal University, Guangzhou 510006, China}
\maketitle

\begin{abstract}
We propose a scheme for generating entangled states of light fields.
This scheme only requires the cross-Kerr nonlinear interaction
between coherent light-beams, followed by a homodyne detection.
Therefore, this scheme is within the reach of current technology. We
study in detail the generation of the entangled states between two
modes, and that among three modes. In addition to the Bell states
between two modes and the W states among three modes, we find
plentiful new kinds of entangled states. Finally, the scheme can be
extend to generate the entangled states among more than three modes.

\end{abstract}

\section{\protect\bigskip Introduction}

Entanglement is a characteristic feature of quantum states and has
important applications in quantum science and technology, for
example, in quantum computation and quantum information
\cite{Nielsen}. There are a lot of schemes for generating various
kinds of entanglement, for example, the entanglement between
photons, the entanglement between atoms, the entanglement between
trapped ions, and the entanglement between different kinds of
particles (for example, between photons and atoms). In addition to
the entanglement between two parties, there are also entanglement of
multiparties. Among these schemes many use single-photon sources
and/or single-photon detectors. Although there are great progresses
in the study on these single-photon devices, how to obtain them is
still a challenging task. In this paper we propose a simple scheme
for generating entangled states of light fields. This scheme only
requires the cross-Kerr nonlinear interaction between light fields
in coherent states, followed by a homodyne detection. Therefore,
this scheme is within the reach of current technology.The basic idea
of this scheme is shown in Figure 1. Mode $a$ is a bright beam which
is in a coherent state $\left\vert \alpha \right\rangle $. Mode $b$
is a weak or bright beam which is also in a coherent state. $BS$ is
a $50/50$ beam splitter. $KM1$ and $KM2$ are Kerr media. $HD$ means
homodyne detection \cite{HD}.

This paper is organized as follows: In section 2 we briefly
introduce the cross-Kerr nonlinear interaction between two
field-modes. In section 3 and section 4 we study the generation of
entanglement between two modes and that among three modes,
respectively. Section 5 is a summary.

\section{Cross-Kerr nonlinear interaction}

First, let us briefly review the cross-Kerr nonlinear interaction between a
mode $A$ and a mode $B.$ The interaction Hamiltonian has the form \cite%
{Sanders}%
\begin{equation}
H_{CK}=\hbar K\hat{n}_{A}\hat{n}_{B},  \label{1}
\end{equation}%
where $\hat{n}_{A}$ and $\hat{n}_{B}$ are the photon-number operator of mode
$A$ and mode $B$, respectively. The coupling coefficient $K$ is proportional
to the third-order nonlinear susceptibility $\chi ^{\left( 3\right) }$. The
time-evolution operator is

\begin{equation}
U\left( t\right) =\exp \left( -\frac{i}{\hbar }H_{CK}t\right) =\exp \left\{
-iK\hat{n}_{A}\hat{n}_{B}t\right\} =\exp \left\{ -i\tau \hat{n}_{A}\hat{n}%
_{B}\right\} =U\left( \tau \right) ,  \label{2}
\end{equation}%
in which $\tau =Kt=K\left( l/v\right) $, it can be named as the scaled
interaction time, or the nonlinear phase shift. Here $l$ is the length of
the Kerr medium and $v$ is the velocity of light in the Kerr medium. The
cross Kerr nonlinearity has following property
\begin{equation}
U\left( \tau \right) \left\vert n\right\rangle _{B}\left\vert \alpha
\right\rangle _{A}=\left\vert n\right\rangle _{B}\left\vert \alpha
e^{-in\tau }\right\rangle _{A},  \label{3}
\end{equation}%
here $\left\vert n\right\rangle $ and $\left\vert \alpha \right\rangle $ are
the photon number state and the coherent state, respectively.

\section{Entanglement between two modes}

Now let us study the generation of the entangled states between two modes.
The scheme is shown in Figure 1. Assume that mode $a$ is in a coherent state
$\left\vert \alpha \right\rangle $ \cite{Gerry2}. Mode $b$ is also in a
coherent state which is divided by the 50/50 beam splitter $BS$ into two
beams $b1$ and $b2$, and both $b1$ and $b2$ are in coherent state $%
\left\vert \beta \right\rangle $.

We \ first consider the case of weak coherent state $\left\vert \beta
\right\rangle $. In this case we have
\begin{equation}
\left\vert \beta \right\rangle \approx \frac{1}{\sqrt{1+\left\vert \beta
\right\vert ^{2}}}\left( \left\vert 0\right\rangle +\beta \left\vert
1\right\rangle \right) ,  \label{4}
\end{equation}%
where $\left\vert 0\right\rangle $ and $\left\vert 1\right\rangle $ are the
vacuum state and one-photon state, respectively. Let mode $a$ interacts with
mode $b1$ and $b2$ successively. For simplicity, we assume that both the
scaled interaction times are $\tau $, thai is, $\tau _{1}=K_{1}t_{1}=\tau
_{2}=K_{2}t_{2}=\tau .$The interactions change the state as following way%
\begin{equation}
\left\vert \beta \right\rangle _{2}\left\vert \beta \right\rangle
_{1}\left\vert \alpha \right\rangle _{a}\rightarrow
\frac{1}{1+\left\vert \beta \right\vert ^{2}}\left[ \left\vert
0\right\rangle _{2}\left\vert 0\right\rangle _{1}\left\vert \alpha
\right\rangle _{a}+\beta \left( \left\vert 1\right\rangle
_{2}\left\vert 0\right\rangle _{1}+\left\vert 0\right\rangle
_{2}\left\vert 1\right\rangle _{1}\right) \left\vert \alpha
e^{-i\tau }\right\rangle _{a}+\beta ^{2}\left\vert 1\right\rangle
_{2}\left\vert 1\right\rangle _{1}\left\vert \alpha e^{-i2\tau
}\right\rangle _{a}\right] ,  \label{5}
\end{equation}%
where the subscripts $1$ and $2$ denote modes $b1$ and $b2$, respectively.
We note that the internal product of coherent states satisfies \cite{Gerry2}%
\begin{equation}
\left\vert \left\langle \alpha e^{-in\tau }|\alpha e^{-i\left( n+1\right)
\tau }\right\rangle \right\vert ^{2}=e^{-4\left\vert \alpha \right\vert
^{2}\sin ^{2}\left( \tau /2\right) }\approx e^{-\left\vert \alpha
\right\vert ^{2}\tau ^{2}},  \label{6}
\end{equation}%
in which we have taken into account the fact that in practice $\tau
$ is small \cite{Sanders} and therefore $\sin \left( \tau /2\right)
\approx \tau /2$. However, if mode $a$ is bright enough so that
$\left\vert \alpha \right\vert ^{2}\tau ^{2}\gg 1,$ then the two
coherent states will be approximately orthogonal. This condition can
be easily satisfied in experiments and in following discussions we
assume that it is satisfied. In this case, a homodyne detection can
distinguish different coherent states \cite{Nemoto}. Therefore, when
we find that mode $a$ is in the coherent state $\left\vert \alpha
e^{-i\tau }\right\rangle _{a}$, then beam $b1$ and beam $b2$ will be
projected into the entangled state
\begin{equation}
\frac{1}{\sqrt{2}}\left( \left\vert 1\right\rangle _{2}\left\vert
0\right\rangle _{1}+\left\vert 0\right\rangle _{2}\left\vert 1\right\rangle
_{1}\right) ,  \label{7}
\end{equation}%
and the probability for getting this entangled state is $2\left\vert \beta
\right\vert ^{2}/\left( 1+\left\vert \beta \right\vert ^{2}\right) ^{2}$
.This state is one of Bell states \cite{Nielsen} and a special case of the $%
NOON$ states \cite{NOON}.

Now let us consider the general situation in which beam $b1$ and beam $b2$
are normal coherent states \cite{Gerry2}. In this situation,%
\begin{equation}
\left\vert \beta \right\rangle =\exp \left( -\frac{1}{2}\left\vert \beta
\right\vert ^{2}\right) \sum\limits_{n=0}^{\infty }\frac{\beta ^{n}}{\sqrt{n!%
}}\left\vert n\right\rangle .  \label{8}
\end{equation}%
The cross-Kerr interactions transform the state as follows%
\begin{eqnarray}
\left\vert \beta \right\rangle _{2}\left\vert \beta \right\rangle
_{1}\left\vert \alpha \right\rangle _{a} &=&e^{-\left\vert \beta \right\vert
^{2}}\sum_{m,n}\frac{\beta ^{m+n}}{\sqrt{m!n!}}\left\vert m\right\rangle
_{2}\left\vert n\right\rangle _{1}\left\vert \alpha \right\rangle _{a}
\notag \\
&\rightarrow &e^{-\left\vert \beta \right\vert ^{2}}\sum_{m,n}\frac{\beta
^{m+n}}{\sqrt{m!n!}}\left\vert m\right\rangle _{2}\left\vert n\right\rangle
_{1}\left\vert \alpha e^{-i\left( m+n\right) \tau }\right\rangle _{a}.
\label{9}
\end{eqnarray}%
If the homedyne detection finds mode $a$ in the state $\left\vert
\alpha e^{-i\left( m+n\right) \tau }\right\rangle _{a}=\left\vert
\alpha e^{-ik\tau }\right\rangle _{a}$ $(k=m+n=1,2,...)$, then mode
$b1$ and mode $b2$ will be collapse
into the entangled state%
\begin{equation}
\frac{1}{\sqrt{2^{k}}}\sum_{n=0}^{k}\sqrt{\frac{k!}{n!\left( k-n\right) !}}%
\left\vert k-n\right\rangle _{2}\left\vert n\right\rangle _{1}\qquad
(k=1,2,...).  \label{10}
\end{equation}%
Since in this state the sum of photon numbers of the two modes is equal to $%
k,$ we name this state as the $2$-mode $k$-photon entangled state.
The probability for getting this state is $exp(-2\left\vert \beta
\right\vert ^{2})\frac{2^{k}}{k!}\left\vert \beta \right\vert
^{2k}.$ The entanglement property of the states expressed by Eq.(10)
can be proved by using following entanglement
criteria \cite{Hillery}%
\begin{equation}
\left\vert \left\langle b_{1}^{+}b_{2}\right\rangle \right\vert
^{2}>\left\langle N_{b1}N_{b2}\right\rangle ,  \label{11}
\end{equation}%
where $N_{b1}(N_{b2}),b_{1}(b_{2})$ and $b_{1}^{+}(b_{2}^{+})$ are
the photon-number operator, the photon annihilation operator and the
photon creation operator of mode $b1(b2)$, respectively. For the
states of equation (10), we can find $\left\vert \left\langle
b_{1}^{+}b_{2}\right\rangle \right\vert
^{2}=\frac{1}{4}k^{2}$ , and $\left\langle N_{b1}N_{b2}\right\rangle =\frac{1%
}{4}k\left( k-1\right) .$ Therefore the entanglement condition (11)
is satisfied, and the states (10) are indeed entangled states. For
$k=1$,
equation (10) reduces to equation (7), and some other examples of the $2$%
-mode $k$-photon entangled states are listed below.%
\begin{equation}
\frac{1}{2}\left[ \left( \left\vert 2\right\rangle _{2}\left\vert
0\right\rangle _{1}+\left\vert 0\right\rangle _{2}\left\vert 2\right\rangle
_{1}\right) +\sqrt{2}\left\vert 1\right\rangle _{2}\left\vert 1\right\rangle
_{1}\right] \qquad (k=2)  \label{12}
\end{equation}

\begin{equation}
\frac{1}{\sqrt{8}}\left[ \left( \left\vert 3\right\rangle _{2}\left\vert
0\right\rangle _{1}+\left\vert 0\right\rangle _{2}\left\vert 3\right\rangle
_{1}\right) +\sqrt{3}\left( \left\vert 2\right\rangle _{2}\left\vert
1\right\rangle _{1}+\left\vert 1\right\rangle _{2}\left\vert 2\right\rangle
_{1}\right) \right] \qquad (k=3)  \label{13}
\end{equation}%
Equations (12) and (13) are new kinds of entangled states. Equation
(12) can be understood as a superposition of a $NOON$ state $\left(
\left\vert 2\right\rangle _{2}\left\vert 0\right\rangle
_{1}+\left\vert 0\right\rangle _{2}\left\vert 2\right\rangle
_{1}\right) $ and a product state $\left\vert 1\right\rangle
_{2}\left\vert 1\right\rangle _{1},$while equation (13) can be
understood as a superposition of a $NOON$ state $\left( \left\vert
3\right\rangle _{2}\left\vert 0\right\rangle _{1}+\left\vert
0\right\rangle _{2}\left\vert 3\right\rangle _{1}\right) $ and a
$NOON-like$ state $\left( \left\vert 2\right\rangle _{2}\left\vert
1\right\rangle _{1}+\left\vert 1\right\rangle _{2}\left\vert
2\right\rangle _{1}\right) $. We also note that in the superposition
(13) the probability of getting the state $\left( \left\vert
2\right\rangle _{2}\left\vert 1\right\rangle _{1}+\left\vert
1\right\rangle _{2}\left\vert 2\right\rangle _{1}\right) $ is larger
than that of getting the state $\left( \left\vert 3\right\rangle
_{2}\left\vert 0\right\rangle _{1}+\left\vert 0\right\rangle
_{2}\left\vert 3\right\rangle _{1}\right) .$That is, the photons
trend to distribute between the two modes symmetrically. The
properties and applications of these new kinds of entangled states
will be studied in the future.

\section{Entanglement among three modes}

We can extend the scheme above to generate the entanglement among
three modes. For this purpose we modify the scheme from Figure 1 to
Figure 2, in which BS1 has the $reflection/transmission=1/2$ and BS2
has the $reflection/transmission=1/1$, so that the three beams
$b1,b2$ and $b3$ have the same strength, and we assume all of them
are in the coherent state $\left\vert \beta \right\rangle .$ We let
mode $a,$ in a coherent state $\left\vert \alpha \right\rangle $,
interacts with modes $b1,b2$ and $b3$ successively. And for
simplicity, we assume that all of the scaled interaction times are
equal, thai is, $\tau _{1}=\tau _{2}=\tau _{3}=\tau .$

For the situation in which $\left\vert \beta \right\rangle $ is weak
and can be expressed as in equation (4), the interactions transform
the states in the following way %
\begin{eqnarray}
\left\vert \beta \right\rangle _{3}\left\vert \beta \right\rangle
_{2}\left\vert \beta \right\rangle _{1}\left\vert \alpha \right\rangle _{a}
&\rightarrow &\frac{1}{\left( 1+\left\vert \beta \right\vert ^{2}\right)
^{3/2}}\{\left\vert 0\right\rangle _{3}\left\vert 0\right\rangle
_{2}\left\vert 0\right\rangle _{1}\left\vert \alpha \right\rangle _{a}
\notag \\
&&+\beta \left( \left\vert 1\right\rangle _{3}\left\vert 0\right\rangle
_{2}\left\vert 0\right\rangle _{1}+\left\vert 0\right\rangle _{3}\left\vert
1\right\rangle _{2}\left\vert 0\right\rangle _{1}+\left\vert 0\right\rangle
_{3}\left\vert 0\right\rangle _{2}\left\vert 1\right\rangle _{1}\right)
\left\vert \alpha e^{-i\tau }\right\rangle _{a}  \notag \\
&&+\beta ^{2}\left( \left\vert 1\right\rangle _{3}\left\vert 1\right\rangle
_{2}\left\vert 0\right\rangle _{1}+\left\vert 1\right\rangle _{3}\left\vert
0\right\rangle _{2}\left\vert 1\right\rangle _{1}+\left\vert 0\right\rangle
_{3}\left\vert 1\right\rangle _{2}\left\vert 1\right\rangle _{1}\right)
\left\vert \alpha e^{-i2\tau }\right\rangle _{a}  \notag \\
&&+\beta ^{3}\left\vert 1\right\rangle _{3}\left\vert 1\right\rangle
_{2}\left\vert 1\right\rangle _{1}\left\vert \alpha e^{-i3\tau
}\right\rangle _{a}\}.  \label{14}
\end{eqnarray}%
As discussed above, we assume that different coherent states in
above equation are approximately orthogonal, and we can use homodyne
detection to distinguish them \cite {Nemoto}. \ If we find that mode
$a$ is in state $\left\vert \alpha e^{-i\tau }\right\rangle _{a}$
then modes $b1,b2$ and $b3$ will be projected
to the entangled state%
\begin{equation}
\frac{1}{\sqrt{3}}\left( \left\vert 1\right\rangle _{3}\left\vert
0\right\rangle _{2}\left\vert 0\right\rangle _{1}+\left\vert 0\right\rangle
_{3}\left\vert 1\right\rangle _{2}\left\vert 0\right\rangle _{1}+\left\vert
0\right\rangle _{3}\left\vert 0\right\rangle _{2}\left\vert 1\right\rangle
_{1}\right) ,  \label{15}
\end{equation}%
and the probability for obtaining this state is $3\left\vert \beta
\right\vert ^{2}/\left( 1+\left\vert \beta \right\vert ^{2}\right) ^{3}$. On
the other hand,\ If we find that mode $a$ is in state $\left\vert \alpha
e^{-i2\tau }\right\rangle _{a}$ then modes $b1,b2$ and $b3$ will be
projected to the entangled state
\begin{equation}
\frac{1}{\sqrt{3}}\left( \left\vert 1\right\rangle _{3}\left\vert
1\right\rangle _{2}\left\vert 0\right\rangle _{1}+\left\vert 1\right\rangle
_{3}\left\vert 0\right\rangle _{2}\left\vert 1\right\rangle _{1}+\left\vert
0\right\rangle _{3}\left\vert 1\right\rangle _{2}\left\vert 1\right\rangle
_{1}\right) ,  \label{16}
\end{equation}%
and the probability for getting this state is $3\left\vert \beta
\right\vert ^{4}/\left( 1+\left\vert \beta \right\vert ^{2}\right)
^{3}$. Equations (15) and (16) can be named as $1$-photon W state
\cite{Dur} and $2$-photon W state, respectively.

For the general case in which $\left\vert \beta \right\rangle $ is not very
weak we use equation (8). In this case the interactions transform the states
as follows:%
\begin{eqnarray}
\left\vert \beta \right\rangle _{3}\left\vert \beta \right\rangle
_{2}\left\vert \beta \right\rangle _{1}\left\vert \alpha \right\rangle _{a}
&=&e^{-3\left\vert \beta \right\vert ^{2}/2}\sum_{l,m,n}\frac{\beta ^{l+m+n}%
}{\sqrt{l!m!n!}}\left\vert l\right\rangle _{3}\left\vert m\right\rangle
_{2}\left\vert n\right\rangle _{1}\left\vert \alpha \right\rangle _{a}
\notag \\
&\rightarrow &e^{-3\left\vert \beta \right\vert ^{2}/2}\sum_{l,m,n}\frac{%
\beta ^{l+m+n}}{\sqrt{l!m!n!}}\left\vert l\right\rangle _{3}\left\vert
m\right\rangle _{2}\left\vert n\right\rangle _{1}\left\vert \alpha
e^{-i\left(l+m+n\right) \tau }\right\rangle _{a}.  \label{17}
\end{eqnarray}

If we find that mode $a$ is in the state $\left\vert \alpha
e^{-i\left( l+m+n\right)
\tau }\right\rangle _{a}=\left\vert \alpha e^{-ik\tau }\right\rangle _{a}$ $%
(k=l+m+n=1,2,...)$, then modes $b1,b2$ and $b3$ will be projected to the
entangled state%
\begin{equation}
\frac{1}{\sqrt{3^{k}}}\sum_{m=0}^{k}\sum_{n=0}^{k-m}\sqrt{\frac{k!}{\left(
k-m-n\right) !m!n!}}\left\vert k-m-n\right\rangle _{3}\left\vert
m\right\rangle _{2}\left\vert n\right\rangle _{1}\qquad (k=1,2,...).
\label{18}
\end{equation}%
We name this state as the $3$-mode $k$-photon entangled state. The
probability for getting this state is $exp(-3\left\vert \beta
\right\vert ^{2})\frac{3^{k}}{k!}\left\vert \beta \right\vert
^{2k}.$The entanglement property of the states of Eq.(18) can be
proved by using following entanglement criteria \cite{Hillery}

\begin{equation}
\left\vert \left\langle b_{1}^{+}b_{2}\right\rangle \right\vert
^{2}>\left\langle N_{b1}N_{b2}\right\rangle \text{ \ and }\left\vert
\left\langle b_{2}^{+}b_{3}\right\rangle \right\vert ^{2}>\left\langle
N_{b2}N_{b3}\right\rangle .  \label{19}
\end{equation}%
For the states (18), we can find $\left\vert \left\langle
b_{1}^{+}b_{2}\right\rangle \right\vert ^{2}=\left\vert \left\langle
b_{2}^{+}b_{3}\right\rangle \right\vert ^{2}=\frac{1}{9}k^{2},$ and $%
\left\langle N_{b1}N_{b2}\right\rangle =\left\langle
N_{b2}N_{b3}\right\rangle =\frac{1}{9}k\left( k-1\right) .$Therefore the
entanglement condition (19) is satisfied, and the states (18) are indeed
entangled states of three modes. For $k=1,$ equation (18) reduces to
equation (15), and some other examples of the $3$-mode $k$-photon entangled
state are as follows:%
\begin{eqnarray}
&&\frac{1}{3}\{\left( \left\vert 2\right\rangle _{3}\left\vert
0\right\rangle _{2}\left\vert 0\right\rangle _{1}+\left\vert 0\right\rangle
_{3}\left\vert 2\right\rangle _{2}\left\vert 0\right\rangle _{1}+\left\vert
0\right\rangle _{3}\left\vert 0\right\rangle _{2}\left\vert 2\right\rangle
_{1}\right)   \notag \\
&&+\sqrt{2}\left( \left\vert 1\right\rangle _{3}\left\vert 1\right\rangle
_{2}\left\vert 0\right\rangle _{1}+\left\vert 1\right\rangle _{3}\left\vert
0\right\rangle _{2}\left\vert 1\right\rangle _{1}+\left\vert 0\right\rangle
_{3}\left\vert 1\right\rangle _{2}\left\vert 1\right\rangle _{1}\right)
\}\qquad \left( k=2\right) ,  \label{20}
\end{eqnarray}%
\begin{eqnarray}
&&\frac{1}{\sqrt{3^{3}}}\{\left( \left\vert 3\right\rangle _{3}\left\vert
0\right\rangle _{2}\left\vert 0\right\rangle _{1}+\left\vert 0\right\rangle
_{3}\left\vert 3\right\rangle _{2}\left\vert 0\right\rangle _{1}+\left\vert
0\right\rangle _{3}\left\vert 0\right\rangle _{2}\left\vert 3\right\rangle
_{1}\right)   \notag \\
&&+\sqrt{3}\left( \left\vert 2\right\rangle _{3}\left\vert 1\right\rangle
_{2}\left\vert 0\right\rangle _{1}+\left\vert 2\right\rangle _{3}\left\vert
0\right\rangle _{2}\left\vert 1\right\rangle _{1}+\left\vert 1\right\rangle
_{3}\left\vert 2\right\rangle _{2}\left\vert 0\right\rangle _{1}+\left\vert
1\right\rangle _{3}\left\vert 0\right\rangle _{2}\left\vert 2\right\rangle
_{1}+\left\vert 0\right\rangle _{3}\left\vert 2\right\rangle _{2}\left\vert
1\right\rangle _{1}+\left\vert 0\right\rangle _{3}\left\vert 1\right\rangle
_{2}\left\vert 2\right\rangle _{1}\right)   \notag \\
&&+\sqrt{6}\left\vert 1\right\rangle _{3}\left\vert 1\right\rangle
_{2}\left\vert 1\right\rangle _{1}\}\qquad \left( k=3\right) .  \label{21}
\end{eqnarray}%
Equation (20) is a superposition of two $2$-photon W states. While equation
(21) is a superposition of a $3$-photon W state (the first line), a product
state (the third line), and a state (the second line) which can be expressed
as%
\begin{equation}
\left\vert 2\right\rangle _{i}\left( \left\vert 1\right\rangle
_{j}\left\vert 0\right\rangle _{k}+\left\vert 0\right\rangle _{j}\left\vert
1\right\rangle _{k}\right) +\left\vert 1\right\rangle _{i}\left( \left\vert
2\right\rangle _{j}\left\vert 0\right\rangle _{k}+\left\vert 0\right\rangle
_{j}\left\vert 2\right\rangle _{k}\right) +\left\vert 0\right\rangle
_{i}\left( \left\vert 2\right\rangle _{j}\left\vert 1\right\rangle
_{k}+\left\vert 1\right\rangle _{j}\left\vert 2\right\rangle _{k}\right) ,
\label{22}
\end{equation}%
where the subscripts $i=1,$or $2,$ or $3,$and $j$,$k$ are the other two,
respectively. We also note that in the superposition (20) the probability of
getting the state $\left( \left\vert 1\right\rangle _{3}\left\vert
1\right\rangle _{2}\left\vert 0\right\rangle _{1}+\left\vert 1\right\rangle
_{3}\left\vert 0\right\rangle _{2}\left\vert 1\right\rangle _{1}+\left\vert
0\right\rangle _{3}\left\vert 1\right\rangle _{2}\left\vert 1\right\rangle
_{1}\right) $ is larger than that of getting the state $\left( \left\vert
2\right\rangle _{3}\left\vert 0\right\rangle _{2}\left\vert 0\right\rangle
_{1}+\left\vert 0\right\rangle _{3}\left\vert 2\right\rangle _{2}\left\vert
0\right\rangle _{1}+\left\vert 0\right\rangle _{3}\left\vert 0\right\rangle
_{2}\left\vert 2\right\rangle _{1}\right) $. This shows again that the
photons trend to distribute among different modes symmetrically.

\section{Summary}
In summary, we have proposed a scheme for generating entangled
states of light fields. This scheme has following advantages: First,
the scheme only involves the cross-Kerr nonlinear interaction
between coherent light-beams, followed by a homodyne detection. It
is not necessary that the cross-Kerr nonlinearity is very large, as
long as the coherent light is bright enough. Therefore, this scheme
is within the reach of current technology. Second, in addition to
the Bell states between two modes and the W states among three
modes, plentiful new kinds of entangled states can be generated with
this scheme. We also found that in the generated entangled states,
the photons have a trend to distribute among different modes
symmetrically. Finally, we would like to point out that the scheme
can be extend to generate the entangled states among more than three
modes.

\textbf{Acknowledgement} This work was supported by  the National
Natural Science Foundation of China under grant nos 60578055 and
10404007.

\textbf{Figure captions}

Figure 1. Scheme for generating entanglement between two modes.
KM:cross-Kerr medium; BS: beam splitter; M: mirror, HD: homodyne
detection.

Figure 2. Scheme for generating entanglement among three modes.
KM:cross-Kerr medium; BS: beam splitter; M: mirror, HD: homodyne
detection.

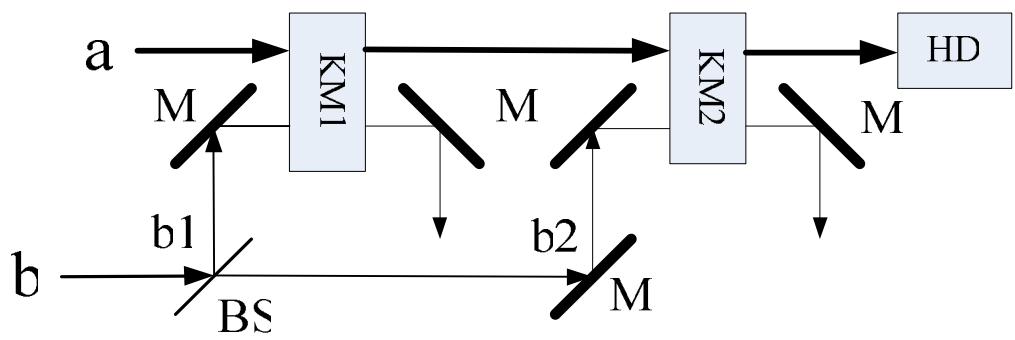
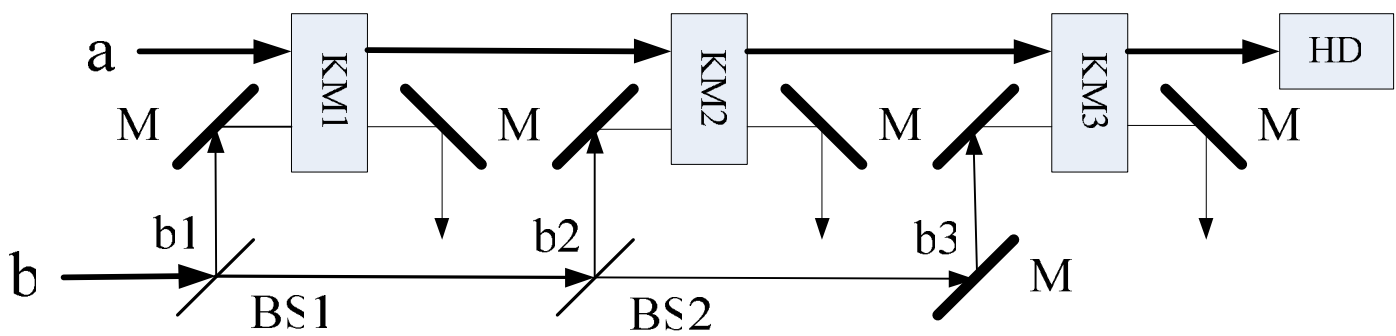
\end{document}